\documentclass[twocolumn,showpacs,preprintnumbers,prl,amsmath,amssymb,superscriptaddress]{revtex4}

\usepackage{graphicx}
\usepackage{endnotes}
\usepackage{bm}
\usepackage{epsfig}

\newcommand{\Co}{CeCoIn$_5$} 
\newcommand{\Rh}{CeRhIn$_5$} 
\newcommand{\RhCo}{CeRh$_{1-x}$Co$_{x}$In$_{5}$}
\newcommand{\ie}{{\it i.e.}} 
 
\newcommand{\etal}{{\it et al.}} 

\begin{document}


\title{Fermi Surface Reconstruction in CeRh$_{1-x}$Co$_{x}$In$_{5}$}


\author{Swee~K.~Goh}
\affiliation{Cavendish Laboratory, University of Cambridge, J.J. Thomson Avenue, Cambridge CB3 0HE, U.K.}

\author{Johnpierre~Paglione}
\affiliation{Department of Physics and Institute for Pure and Applied Physical Sciences, University of California, San Diego, La Jolla, CA 92093}
\affiliation{Center for Nanophysics and Advanced Materials, Department of Physics, University of Maryland, College Park, MD 20742}

\author{Mike~Sutherland}
\author{E.~C.~T.~O'Farrell}
\author{C.~Bergemann}
\affiliation{Cavendish Laboratory, University of Cambridge, J.J. Thomson Avenue, Cambridge CB3 0HE, U.K.}

\author{T.~A.~Sayles}
\affiliation{Department of Physics and Institute for Pure and Applied Physical Sciences, University of California, San Diego, La Jolla, CA 92093}

\author{M.~B.~Maple}
\affiliation{Department of Physics and Institute for Pure and Applied Physical Sciences, University of California, San Diego, La Jolla, CA 92093}

\date{\today}


\begin{abstract}
The evolution of the Fermi surface of CeRh$_{1-x}$Co$_x$In$_5$ was studied as a function of Co concentration $x$ via measurements of the de Haas-van Alphen effect. By measuring the angular dependence of quantum oscillation frequencies, we identify a Fermi surface sheet with $f$-electron character which undergoes an abrupt change in topology as $x$ is varied. Surprisingly, this reconstruction does not occur at the quantum critical concentration $x_c$, where antiferromagnetism is suppressed to $T=0$. Instead we establish that this sudden change occurs well below $x_c$, at the concentration $x\simeq 0.4$ where long range magnetic order alters its character and superconductivity appears. Across all concentrations, the cyclotron effective mass of this sheet does not diverge, suggesting that critical behavior is not exhibited equally on all parts of the Fermi surface.

\end{abstract}


\maketitle


The complex relationship between magnetism and superconductivity in heavy-fermion compounds continues to inspire much experimental and theoretical interest. A key question involves the evolution of the electronic ground state as one tunes from a magnetically ordered state to one in which long range order vanishes. In rare earth compounds, the character of the $4f$ electrons is intimately linked to this question, being dictated by the competition between long range Ruderman-Kittel-Kasuya-Yosida interactions and Kondo screening \cite{Doniach77b}. Varying the magnetic exchange constant $J$ can tune such a system between ground states in which the $4f$ electrons behave very differently -- an antiferromagnetic (AFM) state where they are localized and magnetically coupled via the conduction electrons, and a paramagnetic ground state where the $4f$ moments are subsumed into the conduction electron sea. The character of the $4f$ electrons has direct consequences for the electronic structure, and hence the size and shape of the Fermi surface (FS) \cite{Coleman01}.

In recent years the CeTIn$_5$ (T: Co, Rh, Ir) family has emerged as a fertile testing ground for these ideas, as Ce is a well known Kondo ion and $J$ may be varied in a controlled manner. An extremely rich phase diagram is found by changing T between Co, Rh or Ir: CeCoIn$_5$ and CeIrIn$_5$ are ambient-pressure heavy-fermion superconductors with $T_c=2.3$~K and $0.4$~K, respectively, while CeRhIn$_5$ possesses AFM order below $T_N=3.8$~K \cite{Petrovic_Co,Petrovic_Ir,Hegger00}. Alloying between these compounds \cite{Pagliuso02,Zapf01}, and/or applying pressure \cite{Hegger00,Jeffries05,Park06}, allows for tuning between AFM and superconducting (SC) ground states in a continuous manner. 
In particular, the application of pressure to \Rh\ acts to suppress $T_N$ toward a quantum critical point (QCP) at a critical pressure $P_c = 2.35$ GPa, where a superconducting phase thought to resemble the ambient-pressure state in \Co\ surrounds the QCP \cite{Hegger00}.
The evolution of the FS in \Rh\ through this pressure range was recently investigated by studying the de Haas-van Alphen (dHvA) effect \cite{Shishido05}, which revealed a dramatic FS reconstruction at $P_c$, accompanied by the divergence of the cyclotron mass $m^*$ on several FS sheets. This abrupt FS change was interpreted as the hallmark of a transition from localized to delocalized 4$f$ electron states. The same conclusion was reached in a similar study on CeIn$_3$, a related compound with $T_N$ = 10 K \cite{Settai05}, although this interpretation has been the subject of some debate \cite{Miyake06}.

In this Letter, we present a dHvA study of the electronic structure of \RhCo\ as a function of Co concentration $x$ across the QCP separating the AFM and nonmagnetic phases. Near the critical concentration $x_c$, where $T_N \to 0$, we find no evidence of reconstruction of the observable quasi-2D FS sheet nor any substantial mass enhancement, suggesting that the critical behavior in the CeTIn$_5$ system may be sheet-dependent. Instead, we find a change in FS character to occur \emph{away} from the critical concentration, coincident with a sharp suppression of the superconducting transition temperature and a concomitant change in the magnetic structure.

Previous dHvA studies of CeRhIn$_5$ ($x=0$) \cite{Shishido02b,Shishido05,Hall01} and CeCoIn$_5$ ($x=1$) \cite{Settai01} have established the FS of both compounds to have a common geometry consisting of: (1) a heavy quasi-2D cylindrical sheet giving rise to orbits labeled $\alpha_i$, 
(2) a heavy and more complex quasi-2D sheet with orbits $\beta_i$, and 
(3) several small, light pockets with corresponding low-frequency orbits \cite{Settai01}. 
Band structure calculations \cite{Maehira03} compare well with these observations if the Ce 4$f$ electrons are treated as localized in \Rh\ and itinerant in \Co, a picture further supported by comparisons to the non-$f$-electron analogue LaRhIn$_5$ \cite{Shishido02b,Alver01}.


\begin{figure}[!t]\centering
       \resizebox{8cm}{!}{
              \includegraphics{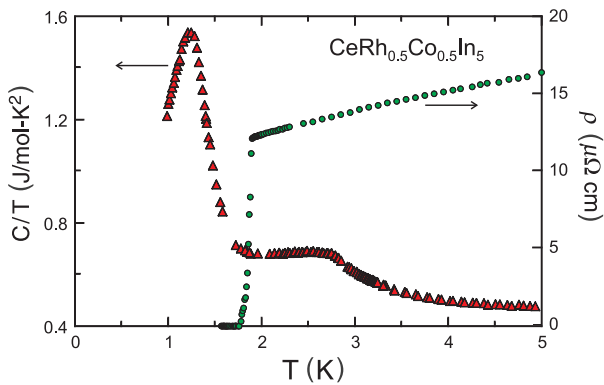}}               				
              \caption{\label{fig1} Temperature dependence of the $a$-axis resistivity (circles) and specific heat (triangles) for \RhCo with Co concentration $x=0.5$, which shows coexistence of antiferromagnetic and superconducting order.}
\end{figure}

Our dHvA studies were performed using high-quality, flux-grown single crystals of \RhCo\ \cite{Jeffries05}. Studies of the $x$-evolution of lattice parameters, resistivity $\rho(T)$, the electronic specific heat $C/T$ and $a$- and $c$-axis DC magnetic susceptibility $\chi(T)$ match the nominal concentrations of these samples with those of previous work \cite{Jeffries05,powder}.
Samples with $x$ $<$~0.4 show signatures only of $T_N$, while samples with 0.4~$<$ $x$~$<$ 0.7 show a coexistence of superconductivity and antiferromagnetism. At higher concentrations, with x $>$~0.7, only $T_c$ is observed. A typical set of data is shown for $x$ = 0.5 in Fig.~1, with $T_N$ and $T_c$ clearly identified by features in both $\rho(T)$ and $C/T$. The observation of dHvA signals from all samples studied suggests that inhomogeneities arising from chemical substitution are small \cite{phasesep}.

Quantum oscillation measurements were carried out to temperatures as low as 10~mK using a second harmonic field modulation technique \cite{Shoenberg84}. We observed oscillations in the field range 14-18~T, and a typical data set is shown in the top inset of Fig.~2. This data is Fourier-transformed to obtain the dHvA frequency $F$ (Fig.~2, main panel), a direct measure of the cross-sectional FS area perpendicular to the applied field. The cyclotron effective mass $m^*$ is extracted from the temperature dependence of the dHvA amplitude by performing a weighted fit of the data with the standard Lifshitz-Kosevich formula \cite{Shoenberg84}, as shown in the bottom inset of Fig.~2. Amplitude uncertainty values are estimated via an average of the integrated background signal for each trace.


\begin{figure}[!t] \centering
  \resizebox{8cm}{!}{
  \includegraphics[width=9cm]{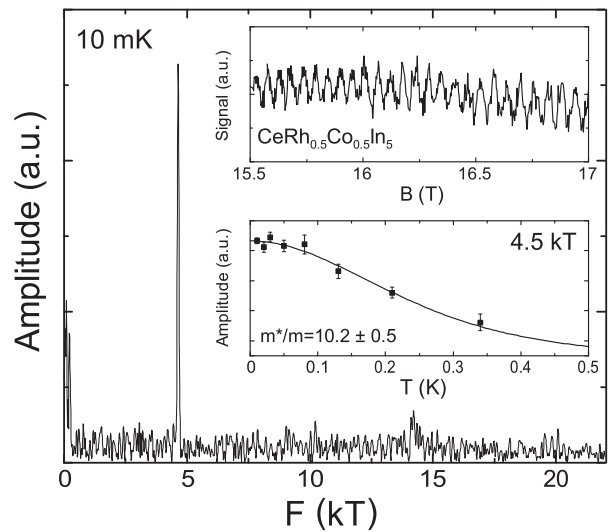}}
  \caption{\label{fig2} Fourier spectrum of quantum oscillations observed in CeRh$_{0.50}$Co$_{0.50}$In$_5$ with $\vec{B}$ $\parallel$ [001], obtained from data collected at 10 mK (upper inset). The temperature dependence of the dHvA amplitude  for the 4.5~kT orbit is fitted (solid line) in the lower inset to extract the effective mass $m^*$.}
\end{figure}


While the introduction of disorder from Rh-Co alloying unavoidably suppresses dHvA amplitudes, direct comparison of oscillation spectra observed at six different $x$ concentrations to those of the end-member compounds makes it possible to extract meaningful FS information throughout the \RhCo\ phase diagram.
In particular, we focus on the rotational dependence of the observed dHvA frequencies $F(\theta)$, obtained by varying the angle $\theta$ between the $c$-axis and the applied field direction $\vec{B}$. Since $F(\theta)$ is uniquely determined by the geometry of the FS, a comparison of $F(\theta)$ for \RhCo\ to previously obtained data on \Rh\ \cite{Shishido02b} and \Co\ \cite{Settai01} allows for an unambiguous determination of exactly where the FS structure changes from `Rh-like' (where the $4f$ electrons are localized) to `Co-like' (where the $4f$ electrons contribute to the FS).

\begin{figure}[!t]\centering
              \resizebox{7.5cm}{!}{
              \includegraphics{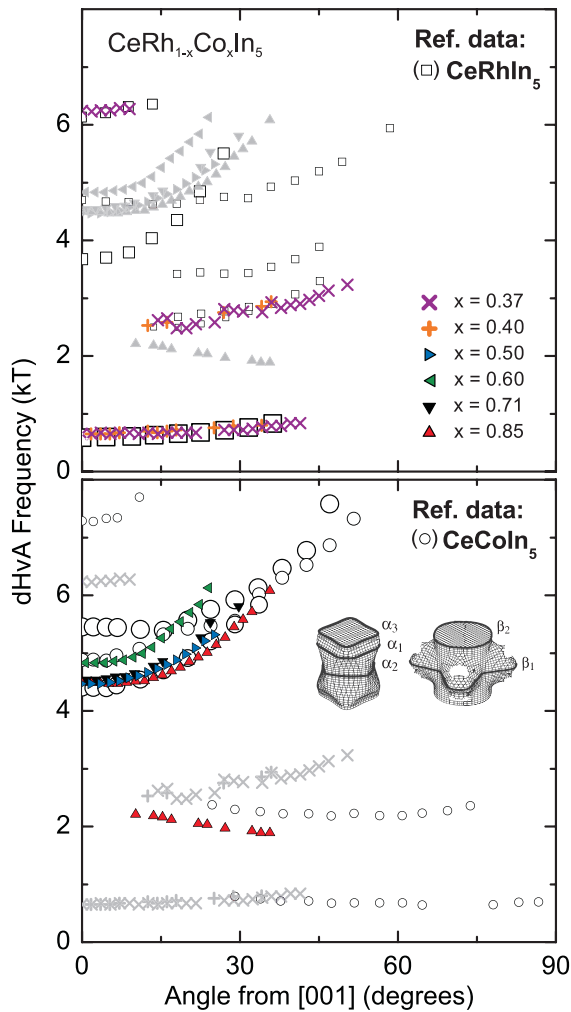}}
              \caption{\label{fig3} (Colour) Rotational dependence of the dHvA frequencies of \RhCo\ samples compared to reference data (open symbols) for both end members of the series: CeRhIn$_5$ ($x=0$) in the top panel \cite{Shishido02b}, and CeCoIn$_5$ ($x=1$)  in the bottom panel \cite{Settai01} (the size of the reference data symbols is normalized to indicate the intensity of a particular orbit in the dHvA spectrum). For all samples we rotate $\vec{B}$ from [001] ($\theta =0$) into the [100] direction. The graphic in the lower panel identifies the observed orbits on the two heavy Fermi surface sheets (adapted from Ref.~\cite{Shishido05}). }
\end{figure}

Fig.~3 presents $F(\theta)$ data for all of our samples, spanning the phase diagram (c.f. Fig.~4). The angular dependence of the observed dHvA frequencies of the $x=0.37$ and $x=0.40$ samples, represented by cross and plus symbols, exhibit excellent agreement with the \Rh\ reference data \cite{alpha1} as shown in the top panel. This is particularly true of the $F(0)=6.2$~kT frequency which follows the $\beta_2$ orbit of \Rh\ (and also that of LaRhIn$_5$) \cite{Shishido02b}.
In contrast, the samples with $x$~$\geq$~0.5 match only the \Co\ reference data, as shown in the lower panel of Fig.~3. In these samples $F(\theta)$ follows the exact angular dependence of the $\alpha_3$ orbit of CeCoIn$_5$ \cite{Settai01}, starting at $F(0)=4.5$~kT and increasing as $\theta$ is increased. 
Hence, it appears that the band structure of \RhCo\ exhibits a change from `Rh-like'  to `Co-like' just above $x=0.40$, deep in the AFM phase and significantly below the critical concentration $x_c$. This is surprising, since the $x=0.50$ sample possesses an AFM ground state (\ie, it lies {\it below} $x_c$), yet the characteristics of its $\alpha$ FS sheet are exactly the same as $x=0.85$ and pure \Co. In fact, as shown in Fig.~4a), the observed heavy $\alpha_3$ orbit seen in {\it all} Co-like samples (\ie, $x>0.40$) does not show any significant change in $F$, and hence its cross-sectional area, as a function of substitution through $x_c$. 

\begin{figure} [!t]\centering
  \resizebox{7.5cm}{!}{
  \includegraphics{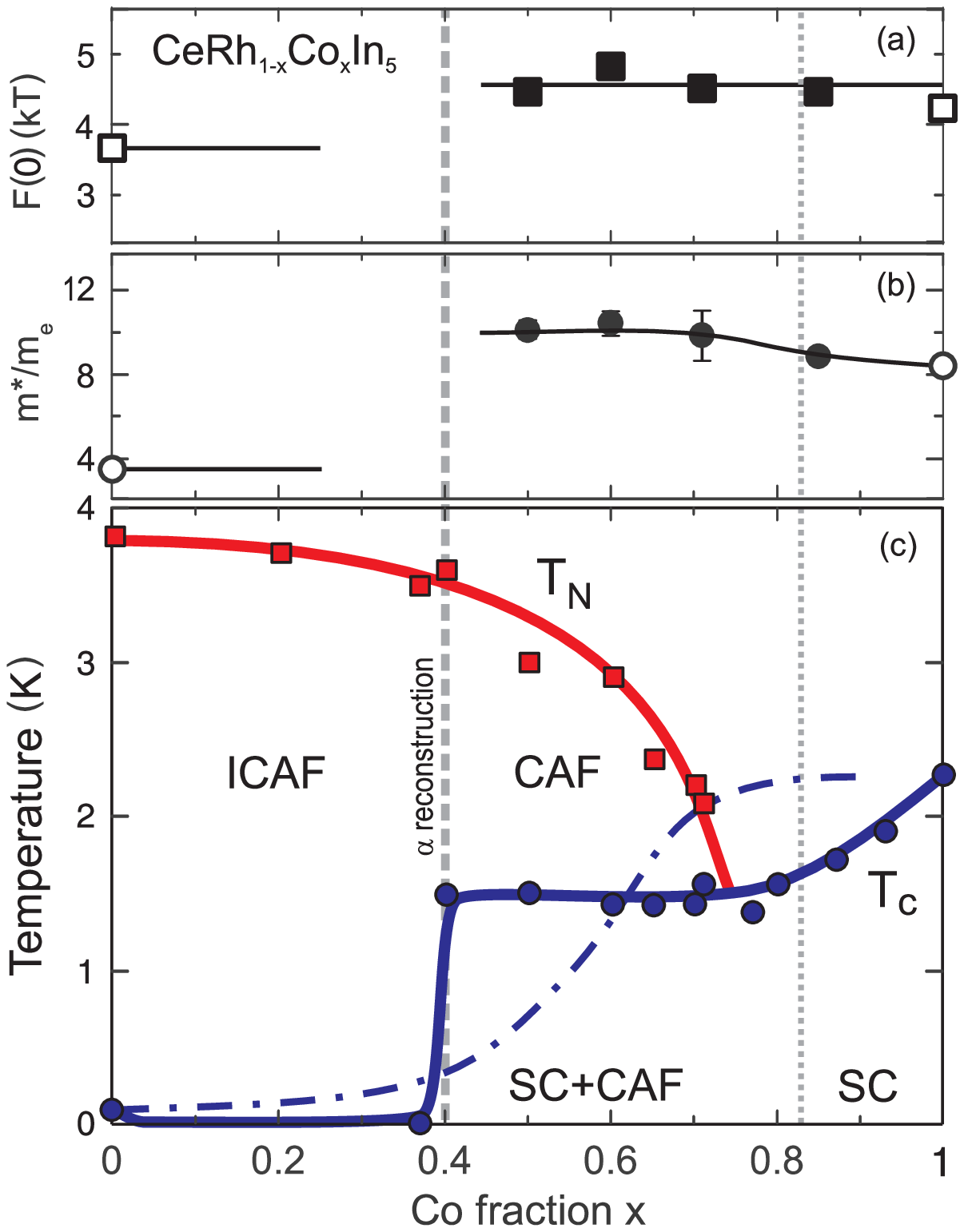}}
  \caption{\label{fig4} Panels (a) and (b) present the dHvA frequency with $\vec{B}$ $\parallel$ [001] and the effective mass evolution, respectively, of the $\alpha_3$ orbit in \RhCo\ as compared to the end members  \Rh\ \cite{Shishido02b} and \Co\ \cite{Settai01} (open symbols), showing minimal change through the antiferromagnetic QCP near $x_c \approx 0.80$ (lines are guides to the eye). Panel (c) maps superconducting (circles) and antiferromagnetic (squares) phase transitions deduced from susceptibility and heat capacity measurements \cite{Jeffries05,Paglione07}, with ``ICAF'' and ``CAF'' denoting incommensurate and commensurate AFM phases, respectively, deduced from neutron scattering \cite{Yokoyama06,Ohira-Kawamura07}. The pressure evolution of $T_c$ (dash-dotted line) and position of the pressure-induced QCP (dotted line) from Ref.~\cite{Chen06}
  are shown for comparison. The pressure axis is linearly scaled to match the concentration and pressure values where $T_N = T_c$ (\ie, at $x\simeq 0.75$ and $P=1.85$~GPa) \cite{Chen06}.
  }
\end{figure}    

This observation lies in contrast to the expectation of an abrupt FS change at a QCP involving a delocalization of $f$-electron states, and is more consistent with a quantum critical spin-density wave scenario where the FS is gradually transformed in a manner determined by the \textbf{q} vector of magnetic fluctuations \cite{Coleman01}. 
However, what's puzzling is that this evolution is empirically different from that obtained in the analogous pressure-tuned system of \Rh, which indeed exhibits a sudden jump in both $\alpha$ and $\beta$ orbits at the critical pressure $P_c=2.35$~GPa \cite{Shishido02b}, where $T_N \rightarrow $ 0 \cite{Park06}. 
This is intriguing, since previous studies have established strong parallels between alloy- and pressure-tuning of these systems \cite{Jeffries05}. In particular, a dramatic increase in the electronic specific heat coefficient centered at $x_c$ suggests an enhancement of $m^*$ at the AFM QCP, which matches the strong enhancement of cyclotron masses observed at $P_c$ for both the $\alpha$ and $\beta$ sheets in \Rh\  \cite{Shishido05}. (In \RhCo, the equivalent critical concentration $x_c$ where $T_N$ $\rightarrow 0$ is around $x \approx 0.8$, as shown by the dotted line in Fig.~4.) As presented in Fig.~4b), our data reveal no attendant divergence of $m^*$ for orbits on the $\alpha$ sheet in this concentration range, despite measuring samples quite close to $x_c$. Taken together with thermodynamic measurements \cite{Jeffries05}, this suggests that critical behaviour may not exist on all sheets equally, indicative of a {\it sheet-} or {\it band-dependent} divergence in the density of states possibly limited to the $\beta$ FS in the \RhCo\ series.

A second distinction between alloy- and pressure-tuning in this system is found in the evolution of $T_c$, which differs in each case. In \Rh, $T_c$ evolves smoothly with increasing pressure from 100~mK at ambient pressure \cite{Paglione07} up to 2~K at $P=1.85$~GPa (where $T_N \simeq T_c$) \cite{Chen06}, with no observed change in FS  \cite{Shishido02b}.
In contrast, superconductivity does not appear \cite{noSC} upon Co substitution until above $x\simeq 0.40$, where it abruptly jumps to $T_c \simeq 1.5$~K  \cite{Jeffries05}. The fact that a significant change in the FS topology of \RhCo\ occurs near this concentration ({\it c.f.} Fig.~3), where superconductivity is strongly enhanced, is provocative.
Theoretical work \cite{Hassan07} has suggested that magnetically mediated spin-singlet superconductivity strongly favors commensurate AFM order. The \RhCo\ system appears to support this idea, since the abrupt onset of $T_c$ coincides precisely with a first-order transition between incommensurate ($[\frac{1}{2},\frac{1}{2},0.297]$) and commensurate ($[\frac{1}{2},\frac{1}{2},\frac{1}{2}]$) AFM order observed in neutron scattering experiments \cite{Yokoyama06,Ohira-Kawamura07}.  
Furthermore, recent neutron scattering measurements of \Co\ have demonstrated a strong coupling between commensurate magnetic fluctuations and superconductivity \cite{Stock07}, which is to be contrasted with the case of \Rh\ where incommensurate  AFM ordering \cite{Llobet04} and superconductivity appear to compete through their entire coexistent pressure range \cite{Paglione07,Chen06}. 

Our observation of a FS reconstruction away from the QCP in \RhCo, coupled with the lack of evidence for a diverging $m^*$ on one heavy FS sheet presents an interesting challenge for leading theories of quantum criticality \cite{Coleman01}. Given the striking similarity of the phase diagram of \RhCo\ to that predicted by recent studies of the Kondo lattice model \cite{Yamamoto07,Watanabe07}, it is tempting to associate the FS change at $x\simeq 0.40$ with a Lifshitz-type transition between AFM states. 
The fact that the FS reconstruction is coincident with a change in the character of magnetic order also suggests that the magnetism may have some itinerant character \cite{Kawasaki03}, whereby a nesting scenario \cite{Ohira-Kawamura07} with a gap opening in the presence of incommensurate ordering may indeed match experimental observations.
While the relationship between superconductivity and the AFM criticality in both pressure- and concentration-tuned CeRhIn$_5$ remains to be determined, the knowledge of the evolution of both the electronic and the magnetic structures contributes significantly to the understanding of the quantum critical behaviour in these two systems.

In summary, quantum oscillation measurements across the concentration-tuned phase diagram of \RhCo\ have uncovered an abrupt change in the Fermi surface that is coincident with a transition between ordered antiferromagnetic phases, occurring well below the quantum critical concentration separating antiferromagnetic and paramagnetic ground states.  The lack of any change of the shape or effective mass associated with the $\alpha$ Fermi surface sheet through this concentration suggests that critical behaviour may not exist on all sheets equally.


The authors acknowledge F. Ronning, M. Ogata, M.A. Tanatar, I. Vekhter and G. Lonzarich for useful discussions. This work is supported by the National Science Foundation (DMR 0335173), Dept. of Energy (DE FG02-04ER56105) EPSRC, the Royal Society, Trinity College and the Yousef Jameel Foundation. J.P. and M.S. acknowledge support from a Natural Science and Engineering Research Council of Canada postdoctoral fellowship and a Royal Society Short Visit grant.



\begin{thebibliography}{34}
\bibitem{powder} X-ray diffraction measurements \cite{Jeffries05} have shown a continuous evolution of lattice parameters as a function of $x$, with both lattice constants increasing linearly as expected by Vegard's law, and the unit cell volume decreasing as expected by the replacement of Rh with Co.
\bibitem{phasesep} The calculated cyclotron length for the largest orbit observed at 6 kT requires grains arising from the presence of any significant phase separation to be at least $0.20~\mu$m at 14~T, highlighting the single-phase nature of our samples.
\bibitem[{alp()}]{alpha1}
\bibinfo{note}{The $\alpha_1$ orbit of CeRhIn$_5$ (also with $F(0)=4.5$~kT) was
  not observed in $x\leq 0.40$ samples likely due to disorder, but the positive
  curvature of the 2.6~kT orbit (Fig.~3) does not match any observed orbits
  in CeCoIn$_5$ \cite{Settai01}.}
\end{thebibliography}
\end{document}